\documentclass[fleqn,usenatbib]{mnras}

\usepackage{newtxtext,newtxmath}

\usepackage[T1]{fontenc}

\DeclareRobustCommand{\VAN}[3]{#2}
\let\VANthebibliography\thebibliography
\def\thebibliography{\DeclareRobustCommand{\VAN}[3]{##3}\VANthebibliography}

\usepackage{graphicx}	
\usepackage{amsmath}	
\usepackage{multirow}    
\usepackage{multicol}    
\usepackage{booktabs}    
\usepackage{tabularx}    
\usepackage{threeparttable}
\usepackage{natbib}
\usepackage{siunitx}
\usepackage[version=4]{mhchem}
\usepackage{hyperref}
\usepackage{xcolor}

\title[GeV $\gamma$-ray emission in the low-mass star-forming region AFGL 490]{ GeV $\gamma$-ray emission in the low-mass star-forming region AFGL 490}

\author[ Yang et al.]{
Li-Nuo Yang,$^{1}$
Sheng Tang,$^{1}$
Pak-Hin Thomas Tam,$^{1,2}$\thanks{E-mail: tanbxuan@sysu.edu.cn }
\\
$^{1}$School of Physics and Astronomy, Sun Yat-sen University, Zhuhai 519082, China\\
$^{2}$CSST Science Center for the Guangdong-Hong Kong-Macau Greater Bay Area, Sun Yat-Sen University, Zhuhai 519082, China\\
}

\date{Accepted XXX. Received YYY; in original form ZZZ}

\pubyear{\the\year{}}

\begin{document}
\label{firstpage}
\pagerange{\pageref{firstpage}--\pageref{lastpage}}
\maketitle

\begin{abstract}
We report the discovery of an extended GeV $\gamma$-ray source, 4FGL~J0330.7+5845e, associated with the star-forming region AFGL 490 using 17 years of \textit{Fermi}-LAT data. The emission is spatially coincident with a dense molecular cloud and centered near the massive protostar AFGL 490. Its spectral energy distribution shows a distinct high-energy cutoff. Both leptonic and hadronic models can fit the $\gamma$-ray spectrum, but energetic arguments rule out stellar winds as the primary accelerator. Instead, the protostellar jet driven by AFGL 490 is identified as a plausible site for particle acceleration, and the derived timescales and maximum particle energies are consistent with theoretical predictions for such jets.
\end{abstract}



\section{Introduction}
\label{intro}
Star-forming (SF) regions, which are often observed to be embedded in giant molecular clouds (GMCs), provide excellent environments for studying particle acceleration. 
The existence of many massive OB stars (sometimes including Wolf-Rayet stars) within a small volume can form young massive clusters (YMC). The intense radiation pressure from the massive stars can drive strong stellar winds with terminal velocity up to $3000$ km s$^{-1}$\citep{1970ApJ...159..879L}. The combined action of stellar winds from multiple massive stars, together with possible supernova remnants within the cluster, can produce a powerful collective cluster wind. The interaction of this collective outflow with the surrounding interstellar medium can generate large-scale shocks and turbulence, providing favorable conditions for particle acceleration ~\citep{2013MNRAS.429.2755B,2021MNRAS.504.6096M}.
YMCs can also sustain high-energy winds over timescales of several million years, fertilizing efficient long-term acceleration of charged particles.

Over the past decade or so, many studies have investigated $\gamma$-ray emission from YMC in SF regions including Cygnus Cocoon~\citep{2011Sci...334.1103A,2021NatAs...5..465A,2024SciBu..69..449L}, NGC 3603~\citep{2017A&A...600A.107Y}, Westerlund 2\citep{2018A&A...611A..77Y}, W43~\citep{2020A&A...640A..60Y}, W40~\citep{2020A&A...639A..80S} and RCW 38~\citep{2024MNRAS.530.1144G}. Typically in the above works, $\gamma$-ray emission can be well explained by the colliding stellar winds model.

However, a less-studied scenario also exists: \citet{2010MmSAI..81..181R} pointed out that, in addition to collective stellar winds from cluster, protostellar jets could also contribute to $\gamma$-ray emission in SF regions. The jets are typically generated during protostellar formation: as the protostar accretes material from the dense GMC in which it is embedded in, it produces jets in which charged particles are accelerated via the Diffusive Shock Acceleration (DSA) mechanism \citep{1978MNRAS.182..147B,1978ApJ...221L..29B}. With velocities ranging from $300$ to $1500$ km s$^{-1}$~\citep{2021MNRAS.504.2405A}, the high-speed jet termination collides with the surrounding dense medium, and $\gamma$ rays are mainly produced through relativistic Bremsstrahlung and proton-proton ($pp$) collisions~\citep{2010A&A...511A...8B}. The latter process generates $\pi^{0}$ mesons (decaying into $\gamma$ rays) as well as charged pions ($\pi^{+}/\pi^{-}$). The subsequent decay of $\pi^{+}/\pi^{-}$ produces secondary electrons/positrons, which can be further accelerated to emit synchrotron radiation. Along this line, \citet{2007A&A...476.1289A,2021MNRAS.504.2405A} and \citet{2010A&A...511A...8B} have developed the model for $\gamma$-ray emission from protostellar jets and placed constraints on the parameters of the jets. Recent studies also suggest that protostars with jets are GeV $\gamma$-ray sources and provide excellent laboratories for investigating particle acceleration mechanisms \citep{2022RAA....22b5016Y,2023MNRAS.523..105D,2025A&A...695A..11M}.

The SF region AFGL 490, as recognized as an embedded cluster of low-mass Young Stellar Objects (YSOs) around the high-mass ($8-10 M_{\odot}$) protostar AFGL 490 \citep{2008hsf1.book..294S,2012ApJ...752..127M}, is located at a distance of approximately $1.0$ kpc~\citep{1984ApJ...284..176S,2008hsf1.book..294S} within the plane of the Galaxy in the Cam OB1 association. In fact, the protostar AFGL 490 resides within a locally over-dense cluster core containing 219 identified YSOs, which are estimated to be around $10^6$ years old with an average mass of about $0.5 M_{\odot}$~\citep{2012ApJ...752..127M}. The protostar AFGL 490 itself has a bolometric luminosity of $2000 L_{\odot}$ \citep{2006ApJ...637L.129S} and an estimated age of $10^4-10^5$ years \citep{2002A&A...394..561S}. Furthermore, multiwavelength observations have revealed its detailed structure, which comprises a prominent bipolar molecular jet \citep{1984ApJ...284..176S,2015MNRAS.450.4364N}, an extended envelope \citep{2002A&A...394..561S}, and a rotating disk \citep{2006ApJ...637L.129S}. The protostar AFGL 490 has also been observed to have two non-thermal radio lobes~\citep{2019MNRAS.486.3664O} similar to HH objects, which confirms the presence of synchrotron radiation, implying the existence of charged particles at relativistic speeds. Based on the position of these non-thermal radio lobes as the termination point of the protostellar jet~\citep{2010A&A...511A...8B}, the jet radius can be inferred to be \(R_j \sim 4.5 \times 10^{16}\) cm.

In this paper, we analysed 17 years of \textit{Fermi}-LAT data to understand the origin of the $\gamma$-ray emission around the SF region AFGL 490. The paper is organized as follows. In Sect.~\ref{data}, we present the data processing procedure for the $\gamma$-ray observations. In Sect.~\ref{gas}, we study the gas distribution in this region. In Sect.~\ref{origin}, we test both leptonic and hadronic models to describe the $\gamma$-ray emission. The results of our discussion are presented in Sect.~\ref{discussion}, and the final conclusions are summarized in Sect.~\ref{conclusion}.

\section{Fermi-LAT DATA ANALYSIS}
\label{data}
The Fermi Large Area Telescope (\textit{Fermi}-LAT) is a space-based gamma-ray observatory capable of surveying a broad energy range from 20 MeV to beyond 1 TeV. The LAT detects $\gamma$-rays by measuring the arrival time, energy, and direction of incident photons using its tracker subsystem and calorimeter.

We performed our analysis using \textit{Fermipy} v1.4.0\footnote{\url{https://fermipy.readthedocs.io/en/latest/}}\citep{2017ICRC...35..824W}. The dataset spans the period from 2008 August 6 (Mission Elapsed Time, MET = 239673601) to 2025 August 6 (MET = 776131205). The \texttt{P8R3\_SOURCE\_V3} instrument response functions (IRFs) were selected for the analysis of events. To maximize the statistical significance, we chose events with \texttt{evtype = 3} and \texttt{evclass = 128}. Additionally, the recommended quality filter expression \texttt{(DATA\_QUAL > 0) \&\& (LAT\_CONFIG == 1)} was applied to select the Good Time Intervals, based on the information provided in the spacecraft file.

Our Region of Interest (ROI) is a $16^{\circ}\times 16^{\circ}$ square centered at the position of  the protostar AFGL 490 with Galactic longitude and latitude ($l=142.00^{\circ}, b=1.82^{\circ}$).
Additionally, to reduce contamination from Galactic diffuse emission, we selected photons in the 300 MeV–300 GeV energy range for our analysis, following, e.g., \citep{2025A&A...695A..11M}.

\subsection{Spatial analysis}
Within the initial $3^\circ \times 3^\circ$ region centered on the protostar AFGL 490, we identified five $\gamma$-ray sources from the \textit{Fermi}-LAT 4FGL Data Release 4 (DR4) catalog~\citep{2022ApJS..260...53A,2023arXiv230712546B}, sorted by their angular offset from the ROI center in ascending order: 4FGL J0330.7+5845 (0.403$^\circ$), 4FGL J0329.2+5750c (0.959$^\circ$), 4FGL J0328.0+6046 (1.988$^\circ$), 4FGL J0323.6+6142 (2.970$^\circ$), and 4FGL J0305.4+5945c (2.993$^\circ$).

\begin{figure}
    \centering
    \includegraphics[width=0.5\textwidth]{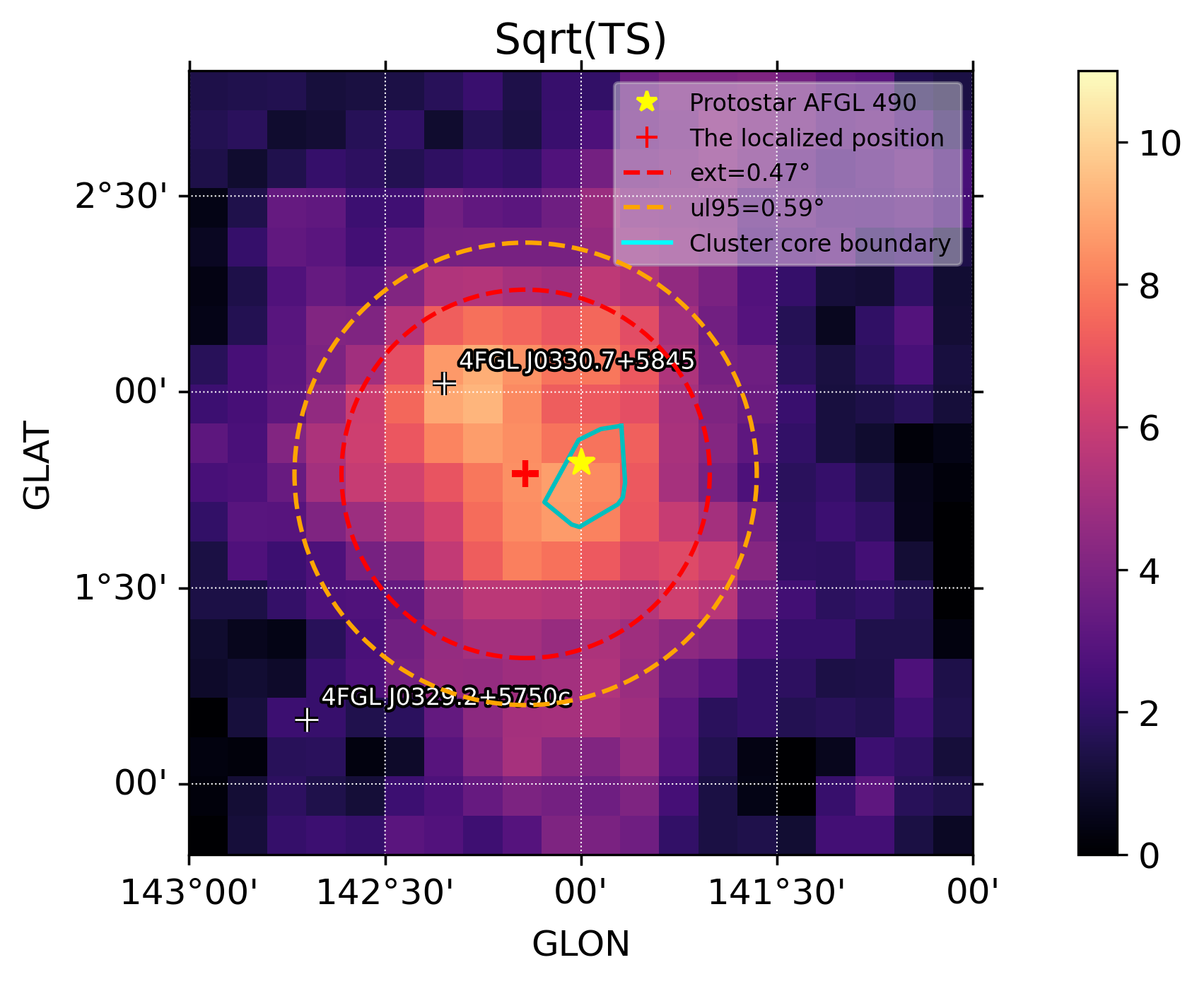}
    \caption{Initial(before \texttt{localize})   Sqrt(TS) map for a $2^\circ \times 2^\circ$ region centered on protostar AFGL 490. Only the source 4FGL J0330.7+5845 was excluded, and the spectral index of the test point source was set to be $2.2$.}  
    \label{TS_map}
\end{figure}

For clarity of presentation, we subsequently only show the $2^\circ \times 2^\circ$ region (see Fig.~\ref{TS_map}). This refined ROI contains only two catalog sources: 
4FGL~J0330.7+5845 ($l = 142.349^\circ$, $b = 2.022^\circ$) 
and 4FGL~J0329.2+5750c ($l = 142.7^\circ$, $b = 1.164^\circ$). In the analysis, all spectral parameters of these two sources are left free during the log-likelihood fitting. In addition, we allowed the normalization parameters of all sources within $5^\circ$ of the ROI center to vary, as well as the normalizations of the Galactic diffuse emission and the isotropic background components.

We conducted a preliminary analysis to delineate the boundaries of the dense core within the AFGL 490 cluster (see Sect.~\ref{intro}) based on the results of \citet{2012ApJ...752..127M}, whose outline is traced by the cyan irregular polygon in Fig.~\ref{TS_map}.

\subsubsection{Localize the source 4FGL J0330.7+5845}
We initially note the apparent presence of a faint dual-peak structure in Fig.~\ref{TS_map}, characterized by one TS peak coincident with the known source 4FGL J0330.7+5845 and another located near the protostar AFGL 490. To investigate this, we added a trial point source with a power-law spectrum at the position of AFGL 490. While the fit converged successfully and produced an increase in the global log-likelihood of $\Delta\log L = 19$, the trial source itself exhibits a TS value of only 7.7 and an anomalously steep spectral index of $\sim$5. Given these characteristics, the feature is unlikely to represent a genuine astrophysical source. Furthermore, a systematic search for new point sources within a $2^\circ \times 2^\circ$ region, conducted using the \texttt{find\_source} method in \textit{Fermipy}, also returned no significant candidates. For a more detailed investigation of this possible two-point-source structure, see Sec.~\ref{revisit_twoPS}.

We then performed a re-localization of the source 4FGL J0330.7+5845 using the \texttt{localize} method in \textit{Fermipy}. The analysis yields an increase of TS from 134.8 to 138.7 and a positional offset of $0.31^\circ$ from the original catalog position. The refined coordinates are ($l=142.14^{\circ}, b=1.79^{\circ}$), which reduces the angular separation from the protostar to $0.144^\circ$. The (re-)localization makes the protostar to lie within the 95\% containment radius ($r_{95} = 0.147^\circ$) of the refined $\gamma$-ray position, which is marked with a red cross in Fig.~\ref{TS_map}.We refer to the model incorporating this updated position as Model~1 in the subsequent analysis.

We also attempted to localize the source 4FGL~J0329.2+5750c (TS = 37.3) after the localization of 4FGL~J0330.7+5845. The localization resulted in a marginal improvement of the fit, with the TS increasing by about 2. The corresponding best-fit position was shifted by only $0.09^\circ$ relative to the catalog value, indicating no significant positional change. We therefore retained its catalog position for the subsequent analysis.

Additionally, we selected photons using the combinedPSF event type, which jointly includes all four PSF event classes, providing the best angular resolution while preserving sufficient statistics, and performed localization for the source 4FGL J0330.7+5845. The result was similar to that obtained with \texttt{evtype = 3}: the position shifted by $0.32^\circ$, with the new position at $(l = 142.14^\circ, b = 1.77^\circ)$. This position is $0.149^\circ$ away from the protostar ($r_{99} = 0.175^\circ$). We adopted this new position and refer to the model after localization with combinedPSF events as Model 2 in the subsequent analysis.

\subsubsection{Extension test for 4FGL~J0330.7+5845}
\label{ext_test}

We employed the \texttt{extension} method in \textit{Fermipy} to test the spatial extension of 4FGL J0330.7+5845 based on Model 1 and Model 2. This analysis was performed across different energy bins and under various assumptions of the source model to ensure robustness. The method determines the best-fit extension parameters by maximizing the likelihood function. Additionally, we conducted a comparative test using photons of the combinedPSF event class only to evaluate its impact on the measured extension.

The Akaike Information Criterion (AIC) \citep{1974ITAC...19..716A} was employed to quantitatively compare and select the better model between the extended source model and the point-source model. AIC is given by the following equation:
\begin{equation}
    \mathrm{AIC} = 2k - 2\ln(L)
\end{equation}
Where $k$ is the number of independently adjusted parameters in the model and $L$ is the maximum value of the likelihood function for the model. The best model is considered to be the one that minimises the AIC value. So, the $\Delta AIC$ = $AIC_{\text{ext}}$ - $AIC_{\text{ps}}$ is used to compare between extension model and initial point source model. In our study, the Gaussian and Disk models were compared against the point-source Model 1, while the Gaussian (combinedPSF) and Disk (combinedPSF) models were compared against the point-source Model 2. Since the spectral shape (logparabola, see Sect.~\ref{sec:spectral}) and spatial position were fixed, switching from a point source to an extended source only increases the number of free parameters $k$ by 1. Additionally, the $ln(L)$ values can be directly obtained from the \textit{Fermipy} output.

The results of these analyses are summarized in Tab.~\ref{ext table}, where the $\Delta \mathrm{AIC}$ values were computed using the full energy range of 0.3–300 GeV. To investigate whether the observed spatial extension exhibits an energy dependence, the analysis was additionally performed in two separate energy intervals: 0.3–1 GeV and 1–300 GeV, with both models providing strong and consistent evidence in favor of spatial extension.

\renewcommand{\arraystretch}{1.2}
\begin{table*}
\centering
\caption{Results of the extension test for 4FGL J0330.7+5845. The source has TS value of 138.7 and 132.7 (combinedPSF) within the 300 MeV--300 GeV energy range. The extension radius and the corresponding 95\% confidence level upper limits (UL95) for the four spatial models in the 300 MeV--300 GeV energy range are also reported. $TS_{\rm ext}$ denotes the test statistic for the spatial extension hypothesis. A $|\Delta \mathrm{AIC}| > 5$ provides strong evidence in favor of the model with the lower AIC, while a $|\Delta \mathrm{AIC}| > 10$ offers conclusive evidence \citep{10.1111/j.1745-3933.2007.00306.x,2017ApSS.362...70K,2025SCPMA..6879503L}.}
\label{ext table}
\begin{tabular}{lccccc}
\hline
Model 
& Extension radius / UL95 ($^\circ$)
& $\Delta$AIC 
& $TS_{\rm ext}(0.3\text{--}300~\mathrm{GeV})$ 
& $TS_{\rm ext}(0.3\text{--}1~\mathrm{GeV})$ 
& $TS_{\rm ext}(1\text{--}300~\mathrm{GeV})$ \\
\hline
Gaussian          & $0.47^{+0.08}_{-0.06} / 0.59$ & -31.0 & 32.8 & 22.9 & 74.2 \\
Disk              & $0.45^{+0.05}_{-0.05} / 0.54$ & -30.1 & 32.0 & 20.4 & 68.8 \\
Gaussian (combinedPSF)   & $0.46^{+0.07}_{-0.06} / 0.58$ & -36.3 & 38.5 & 27.3 & 87.0 \\
Disk (combinedPSF)       & $0.46^{+0.05}_{-0.05} / 0.54$ & -35.1 & 37.3 & 24.8 & 81.0 \\
\hline
\end{tabular}
\end{table*}

\subsubsection{Re-examining the possible two-point-source structure}
\label{revisit_twoPS}
We performed additional tests to investigate the possible dual-peak structure. First, we selected photons with energies $>1$ GeV, which provide a significantly improved PSF. The corresponding TS map still shows a similar two-peak morphology as in Fig.~\ref{TS_map}. Using this data set, we applied the \texttt{find\_sources} method in \texttt{fermipy}, adopting an assumed spectral index of 2.2, a TS detection threshold of 16, and a minimum separation of $0.1^\circ$. However, no additional source was detected in the vicinity (the nearest newly detected source lies more than $2.6^\circ$ away). 

We then added a point source with a power-law spectrum at the location of the second peak ($l = 142.1^\circ$, $b = 1.6^\circ$) and performed a likelihood fit. Before adding this source, the TS value of 4FGL~J0330.7+5845 was 66.3; after the fit, the TS values of the newly added source and 4FGL~J0330.7+5845 become 39.6 and 40.8, respectively. The resulting spectral indices of both sources (with 4FGL~J0330.7+5845 modeled by a log-parabola) are $\sim 3.2$, indicating similarly soft spectra. The comparable TS values and nearly identical soft spectral indices suggest that the emission is likely produced by the same underlying mechanism, and that the fit is likely redistributing the same photon population rather than revealing two independent emitters. 

In addition, we performed a dedicated localization and extension analysis of 4FGL~J0330.7+5845 using only photons with energies $>1$ GeV, following the same procedure as described above. The best-fit position is found to be $ (l=142.13^\circ, b=1.73^\circ)$, with an offset of $0.159^\circ$ from the protostar AFGL~490 (with $r_{99} = 0.195^\circ$), and a TS value of 74.1. At this location, an extension test was carried out using a Gaussian spatial template, yielding $\mathrm{TS}_{\mathrm{ext}} = 31.4$, with a best-fit extension radius of $0.47^{+0.07}_{-0.06}{}^\circ$ and a 95\% upper limit of $0.59^\circ$. Furthermore, compared with the manually constructed two-point-source model, the extended-source model is preferred with $\Delta \mathrm{AIC} = -5.7$. Combined with the consistent evidence for significant extension across all energy bands (see Tab.~\ref{ext table}), these results further support a single extended-source interpretation. 

On the other hand, the photon energies in this region are only at the level of a few GeV (see Fig.~\ref{brems+pp}), at which the angular resolution of the \textit{Fermi}-LAT remains limited (e.g., $\sim0.8^\circ$ at $\sim1$ GeV), and is therefore insufficient to reliably resolve such a marginal and closely spaced ($\sim0.5^\circ$) double-source structure. In light of this limitation, together with the evidence presented above, we adopt a single extended-source model for 4FGL~J0330.7+5845 in the subsequent analysis, using a Gaussian spatial template as the final model and designating the source as 4FGL~J0330.7+5845e. The subsequent analysis was therefore performed using the extension radius of 4FGL~J0330.7+5845e derived from the 0.3–300 GeV energy band, which was fitted as $0.47^\circ$ with a 95\% upper limit of \qty{0.59}{\degree}. Both of these radii are indicated in Fig.~\ref{TS_map}.

\subsection{Spectral energy distribution}
\label{sec:spectral}
We initially used the log-parabolic (LP) spectral model of the original point source 4FGL J0330.7+5845 for the source 4FGL~J0330.7+5845e, which is expressed as:
\begin{equation}
\frac{dN}{dE} = N_0 \left( \frac{E}{E_b} \right)^{-\alpha - \beta \ln(E / E_b)}
\end{equation}
 Our spectral fitting yielded the following parameter values: $N_0 = (1.847 \pm 0.170) \times 10^{-12}\ \mathrm{MeV^{-1}\ cm^{-2}\ s^{-1}}$, $\alpha = 2.312 \pm 0.112$, $\beta = 0.234 \pm 0.089$, and $E_b = 1038$ MeV. Then we tested the spectral curvature of 4FGL~J0330.7+5845e using the \texttt{curvature} method in \textit{Fermipy}. The results show that both curved spectral models provide a better fit than a simple power-law model. Specifically, we obtain a curvature test statistic of 8.4 for the LP model and 9.9 for the PLSuperExpCutoff4\footnote{\url{https://fermi.gsfc.nasa.gov/ssc/data/analysis/scitools/source_models.html}}(PLSC), with the latter computed by fixing the high-energy index to 0.667. The difference in fit quality between the LP and PLSC models is not statistically significant; therefore, we retained the LP model for the subsequent analysis. The derived spectrum reveals a significant high-energy turn-over feature.

\section{GAS CONTENT}
\label{gas}
We investigated three distinct gas phases---molecular hydrogen (\ce{H2}), neutral atomic hydrogen (\ce{HI}), and ionized hydrogen (\ce{HII})---in the vicinity of the star-forming region AFGL 490.

\subsection{The \ce{H2} distribution}
\begin{figure}
    \centering
    \includegraphics[width=0.5\textwidth]{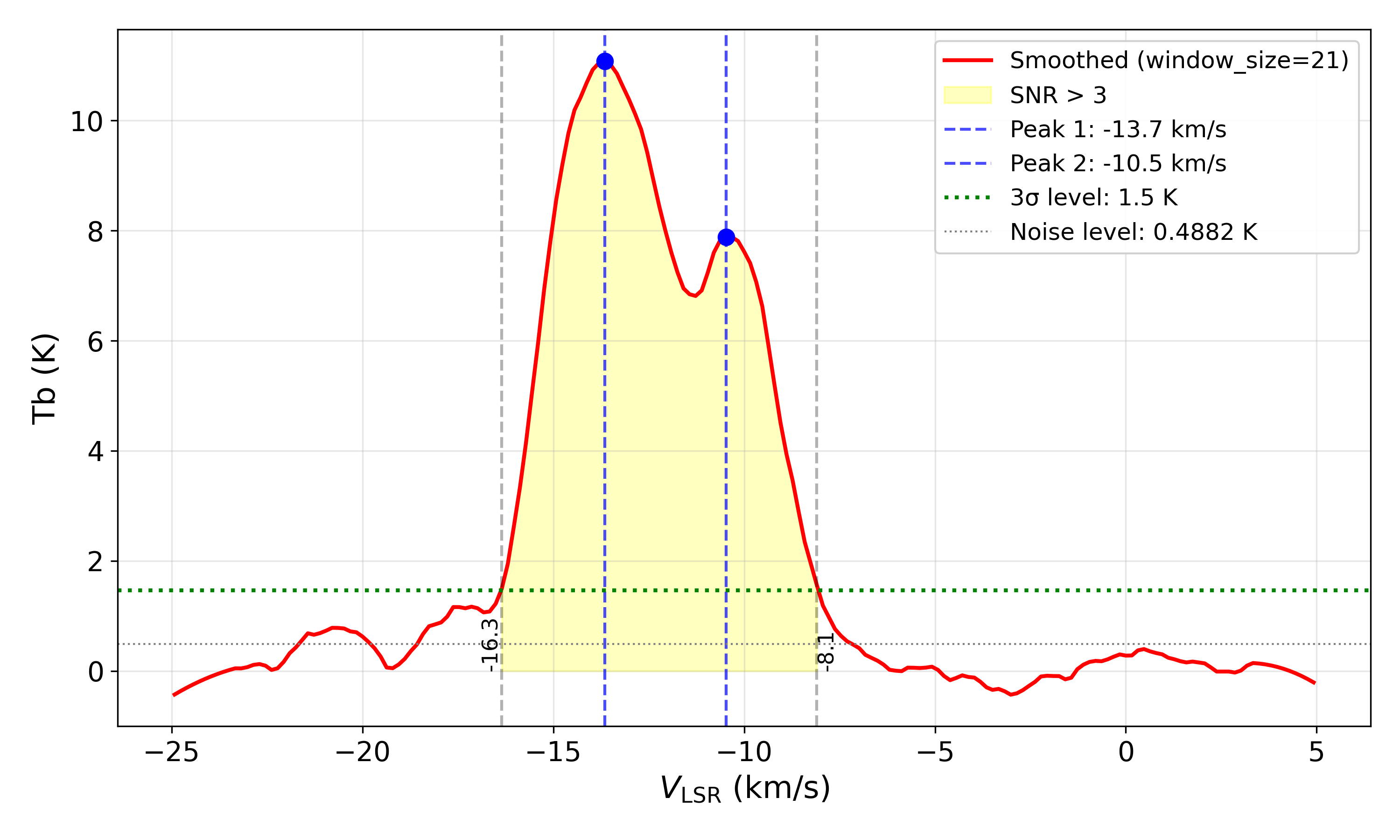}
    \caption{The Savitzky-Golay–smoothed $^{12}$CO spectrum at the position of the molecular cloud PGCC G142.03+1.73, with the noise level indicated and the regions with SNR > 3 marked.}  
    \label{COline}
\end{figure}

\begin{figure}
    \centering
    \includegraphics[width=0.5\textwidth]{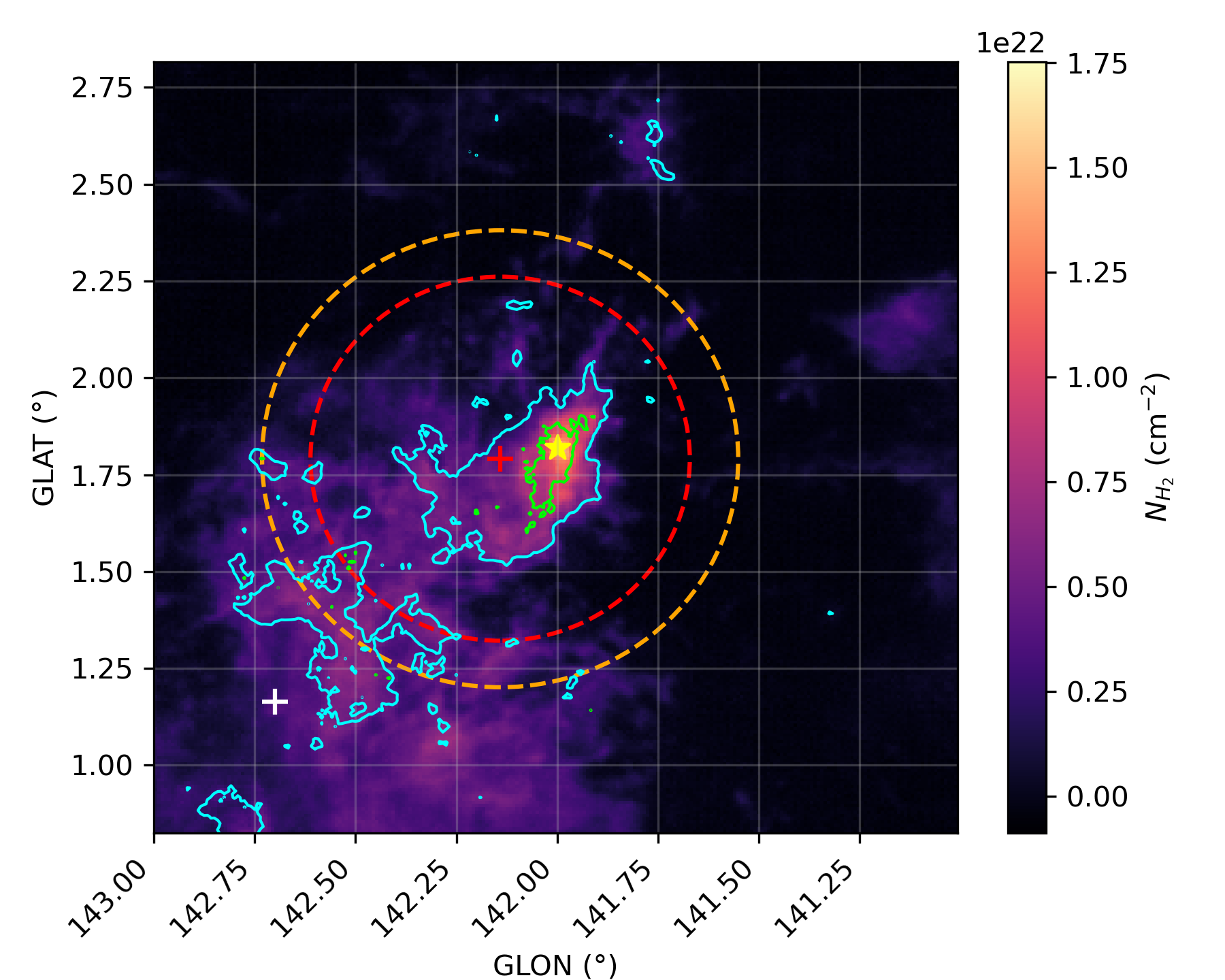}
    \caption{Map of the molecular hydrogen column density within the central $2^\circ \times 2^\circ$ region of interest (ROI). The $^{13}$CO emission contours are outlined in cyan, corresponding to 10\% of the peak intensity ($\sim 3.97$ K km/s), and the C$^{18}$O emission contours are outlined in lime, corresponding to 10\% of the peak intensity ($\sim 0.58$ K km/s). The red cross marks the source center of 4FGL~J0330.7+5845e, together with its extension radius (red, dashed circle) and $ul_{95}$ (orange, dashed circle). The yellow star denotes the protostar AFGL 490, while the white cross marks the source 4FGL J0329.2+5750c. } 
    \label{Hcontours}
\end{figure}

We utilized $^{12}$CO/$^{13}$CO/C$^{18}$O observational data from the MWISP survey~\citep{2019ApJS..240....9S} covering a 4°×4° region centered on the ROI to trace the molecular hydrogen distribution. Through queries of the SIMBAD astronomical database~\citep{2000A&AS..143....9W} and consideration of cluster distances, we selected the molecular cloud PGCC G142.03+1.73 $(l = 142.034^\circ, b = 1.728^\circ)$ \citep{2016A&A...594A..28P} for CO spectral analysis, as it represents the cloud closest to the protostar AFGL 490 and most likely to be spatially associated with the cluster. To identify velocity peaks, we applied Savitzky-Golay filtering~\citep{1964AnaCh..36.1627S} to smooth the initial $^{12}$CO spectra, revealing two distinct velocity components (see Fig.~\ref{COline}): Peak 1 at $-13.7$ km s$^{-1}$ (consistent with the reported  $V_{\rm LSR}$ = $-13.4$ km s$^{-1}$ at the position of protostar AFGL 490~\citep{2006ApJ...637L.129S}) and Peak 2 at $-10.5$ km s$^{-1}$. Using the Galactic measurements from \citet{2019ApJ...885..131R}, we derived kinematic distances of 1.05 kpc and 0.79 kpc for these two peaks, respectively. We subsequently integrated only Peak 1, which is consistent with the cluster distance, over the velocity range of $[-16.3, -11.27]$ km s$^{-1}$ where the Signal-to-Noise Ratio (SNR) $>$ 3. Adopting a conversion factor of $X_{\mathrm{CO}} = 2\times10^{20}\ \mathrm{cm^{-2}\ K^{-1}\ km^{-1}} $ s suggested by \citet{2001ApJ...547..792D} and \citet{2013ARA&A..51..207B}, we obtained the $N_{\text{H}_2}$ distribution map derived from integration of the spectral data of $^{12}$CO (see Fig.~\ref{Hcontours}). Additionally, we integrated the $^{13}$CO spectra (peak velocity: $-12.8$ km s$^{-1}$, integration range: $[-15.4, -9.6]$ km s$^{-1}$ with SNR > 3) and the C$^{18}$O spectra (peak velocity: $-12.8$ km s$^{-1}$, integration range: $[-14.1, -11.6]$ km s$^{-1}$ with SNR > 3) to generate their respective intensity distribution maps. We then delineated the $^{13}$CO and C$^{18}$O intensity contours at a threshold of 10\% of their peak intensities (see Fig.~\ref{Hcontours}). Our analysis reveals a good spatial correspondence between the integrated emission of Peak 1 and both the $^{13}$CO and C$^{18}$O contours. Therefore, we adopted the integration results of Peak 1 for subsequent studies.

\subsection{The \ce{HI} distribution}
We investigated the neutral hydrogen distribution using the HI4PI~\citep{2016A&A...594A.116H} 21-cm data cube. To impose the same kinematic constraint as for Peak 1, the \ce{HI} spectra were integrated over the identical velocity interval $[-16.3,,-11.27]\ \mathrm{km\ s^{-1}}$. We found the \ce{HI} brightness temperature reaches $\sim 80\ \mathrm{K}$ at this interval, indicating that the emission is not optically thin~\citep{1990ARA&A..28..215D}, so we computed the \ce{HI} column density using the following equation:
\begin{equation}
N_{\mathrm{HI}} = -\,1.823 \times 10^{18}\, T_s
\int \ln \left[ 1 - \frac{T_B}{T_s - T_{\mathrm{bg}}} \right] \, {\rm d}v 
\end{equation}
where $T_{\mathrm{bg}} \approx 2.66\ \mathrm{K}$ is the 21-cm CMB brightness temperature, and $T_B$ denotes the bright temperature of the \ce{HI} emission. When $T_B > T_s - 5\ \mathrm{K}$, it is truncated to this value, with the spin temperature fixed at $T_s = 150\ \mathrm{K}$. From this, we obtained a map of the neutral hydrogen column density within a $2^\circ \times 2^\circ$ region centered on the ROI
and found that the distribution of \ce{HI} exhibits a diffuse background, showing no significant spatial correlation with the \ce{H2} distribution.

\subsection{The \ce{HII} distribution}
\label{HII}
We used the Planck free-free map~\citep{2016A&A...594A..10P} to obtain the \ce{HII} density. First, we converted the emission measure (EM) into the free-free intensity using the conversion factor provided in Table 1 of \citet{2003ApJS..146..407F}. We then derived the \ce{HII} column density from the free-free intensity ($I_\nu$) following Eq. (5) of \citet{1997ApJ...480..173S}):
\begin{equation} 
\begin{aligned}
N_{\mathrm{H\,II}} &= 1.2 \times 10^{15}\ \mathrm{cm^{-2}}
\left( \frac{T_e}{1\ \mathrm{K}} \right)^{0.35}
\left( \frac{\nu}{1\ \mathrm{GHz}} \right)^{0.1} \\
&\quad \times \left( \frac{n_e}{1\ \mathrm{cm^{-3}}} \right)^{-1}
\frac{I_\nu}{1\ \mathrm{Jy\ sr^{-1}}} \ ,
\end{aligned}
\end{equation}
where $\nu = 353~\mathrm{GHz}$ is the frequency, and the electron temperature is 
$T_e = 8000~\mathrm{K}$. We then adopt an effective electron density of $2~\mathrm{cm^{-3}}$, as suggested by \citet{1997ApJ...480..173S} for the region outside the solar circle, to obtain a map of the ionized hydrogen column density within a $2^\circ \times 2^\circ$ region centered on the ROI. We find that the distribution of ionized hydrogen also exhibits a diffuse morphology and shows no significant spatial correlation with the molecular hydrogen distribution. On the other hand, \citet{1983ApJ...266..623S} previously estimated that an ultra-compact \ce{HII} region exists at the center of the protostar AFGL 490, but its size is very small ($\lesssim 0.1''$) and therefore does not produce any detectable structure at our observational resolution.

\subsection{Estimating total proton number density} 
\label{density}
Firstly, we calculate the total proton mass within the extension radius $ext = 0.47^\circ$ of source 4FGL~J0330.7+5845e by integrating the column densities of three hydrogen species over the source region. The mass of each component is obtained by multiplying its column density by the angular area of the source 4FGL~J0330.7+5845e, the square of the distance ($d = 1$ kpc), and the respective mass per particle. The total proton mass is then given by \(
M_{\rm P} = M_{\ce{H2}} + M_{\ce{HI}} + M_{\ce{HII}},\)
which represents the total mass of the target material. The derived masses for each component, along with the total mass, are summarized in Tab.~\ref{mass}.

\begin{table}
\centering
\caption{The masses for each component, along with the total mass inside the extension radius $ext = 0.47^\circ$ of source 4FGL~J0330.7+5845e.}
\label{mass}
\begin{tabular}{lc}
\hline
Component & Mass \\
\hline
\ce{H2} & $8.45 \times 10^{3} M_{\odot}$ \\
\ce{HI} & $1.71 \times 10^{3} M_{\odot}$ \\
\ce{HII}  &  $7.14 \times 10^{2} M_{\odot}$\\
\ce{H2} + \ce{HI} + \ce{HII} & $1.09 \times 10^{4} M_{\odot}$ \\
\hline
\end{tabular}
\end{table}

Next, we estimate the size of the molecular cloud within the extension radius using the formula $R =  \sqrt{A / \pi}$, where $A$ is the total area of pixels with a molecular hydrogen column density exceeding $1 \times 10^{21} \, \mathrm{cm^{-2}}$. This yields an estimated radius of $R \sim 6.5 \, \mathrm{pc}$. With this radius, we model the cloud portion as a uniform sphere, i.e., $V = \frac{4}{3}\pi R^3$, and derive an average proton number density of $n_p \sim 375 \, \mathrm{cm^{-3}}$.

\section{THE ORIGIN OF $\gamma$-RAY EMISSION}
\label{origin}
To investigate the $\gamma$-ray radiation mechanisms in the star-forming region AFGL 490, we fitted the spectral energy distribution (SED) of 4FGL~J0330.7+5845e with both leptonic and hadronic models using \textsc{Naima} \citep{2015ICRC...34..922Z}. This computational package provides tools for modeling non-thermal radiation and enables Markov Chain Monte Carlo (MCMC)-based fitting to observational data.

\subsection{Leptonic scenario}
We adopt $n \sim 375  \mathrm{cm^{-3}}$ (see Sect.~\ref{density}) as the number density of the target material for relativistic Bremsstrahlung, and then use the following formula~\citep{2004vhec.book.....A} to estimate the cooling timescale:
\begin{equation}
t_{\mathrm{br}} = \frac{\varepsilon_{e}}{-\,\mathrm{d}\varepsilon_{e}/\mathrm{d}t}
\simeq 4 \times 10^{7} \left( \frac{n}{1~\mathrm{cm^{-3}}} \right)^{-1}~\mathrm{yr}.
\end{equation}
This yields a value of 
$t_{\mathrm{br}} \simeq 1.07\times 10^{5}~\mathrm{yr} $. Furthermore, we tested three electron distribution spectra for relativistic Bremsstrahlung: Power Law (PL), Log-Parabola (LP), and Exponential Cutoff Power Law (ECPL). The Bayesian Information Criterion (BIC) was used for the selection of models~\citep{1978AnSta...6..461S}, with the results presented in Tab.~\ref{diffrent model}. Ultimately, the ECPL spectrum was adopted for the electron distribution:
\begin{equation}
\label{ECPL}
f(E) = A \left( \frac{E}{E_0} \right)^{-\alpha} \exp\left( -\frac{E}{E_{\text{cutoff}}} \right)
\end{equation}
where $E_0 = 1.7 \ \text{GeV}$, the electron spectral index, $\alpha = 1.6_{-0.4}^{+0.4}$, and
$E_{\text{cutoff}} = 3.7_{-1.2}^{+2.6} \ \text{GeV}$. The total electron energy above $3 \ \text{GeV}$ is $W_e({>}3\ \text{GeV}) = 8.1_{-1.3}^{+1.3} \times 10^{44} \ \text{erg}$, and above $0.3 \ \text{GeV}$ it is $W_e({>}0.3\ \text{GeV}) = 3.1_{-0.3}^{+0.3} \times 10^{45} \ \text{erg}$; the corresponding maximum log-likelihood value for this fit is $-0.51$. The relatively large uncertainties in the spectral index and $E_{\rm cutoff}$ are likely due to the small number of high-energy photons, which limits the MCMC's ability to constrain these parameters. In addition, although \citet{2019MNRAS.486.3664O} detected non-thermal radio lobes at L-band, the non-detection at C-band means that only an upper limit on the spectral index ($\alpha_{\rm LC} < -0.51$ and $< -0.63$, respectively) could be inferred. Its usefulness in our modeling is not clear, because our fit is primarily driven by the $\gamma$-ray data, and it remains unclear whether the electron population responsible for the non-thermal radio emission is the same (either spatially or energetically) as that produces the $\gamma$-ray emission. As a result, the available radio constraint cannot be straightforwardly incorporated to significantly improve the parameter determination. The fitted SED result for relativistic Bremsstrahlung of source 4FGL~J0330.7+5845e is shown in Fig.~\ref{brems+pp}.

On the other hand, considering the dense molecular cloud environment within the extension radius of 4FGL~J0330.7+5845e and the lack of massive OB stars in this region (see Sect.~\ref{intro}), the UV radiation field is weak, leading to a low photon energy density. Therefore, the efficiency of inverse Compton scattering is expected to be relatively low. Consequently, we disregard the inverse Compton process in our modeling.

\subsection{Hadronic scenario}
\label{hadronic}
As shown in Fig.~\ref{Hcontours}, the $\gamma$-ray emission from 4FGL~J0330.7+5845e largely encompasses the dense regions of the molecular cloud, as traced by \ce{^{13}CO} and \ce{^{18}CO}. This spatial correspondence provides strong support for a hadronic origin, as the cloud supplies abundant target material.
We also estimate the cooling time~$t_{pp}$ of relativistic protons due to inelastic~$pp$ interactions by using the following formula~\citep{2004vhec.book.....A}:
\begin{equation}
t_{\mathrm{pp}} = \left( n_{0}\,\sigma_{\mathrm{pp}}\,f\,c \right)^{-1} 
\simeq 5.3 \times 10^{7}\,\left( \frac{n}{1~\mathrm{cm^{-3}}} \right)^{-1}~\mathrm{yr},
\end{equation}
where we again adopt $n \sim 375\,\mathrm{cm}^{-3}$ (see Sect.~\ref{density}) and have finally obtained $t_{pp} \simeq 1.41 \times 10^{5} \ \mathrm{yr} $.

\begin{figure}
    \centering
    \includegraphics[width=0.5\textwidth]{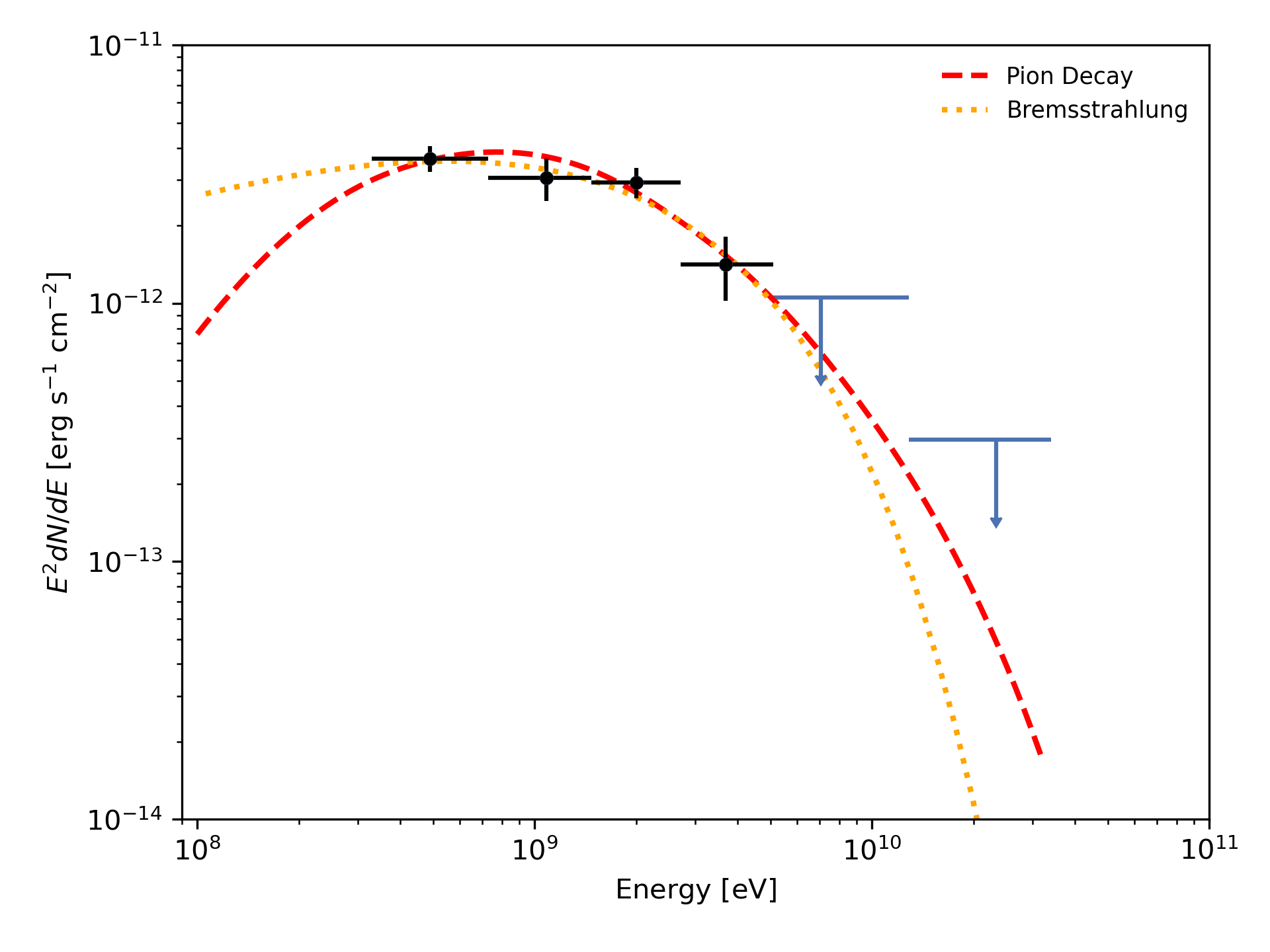}
    \caption{The spectral energy distribution (SED) resulting from the $pp$ interaction process and relativistic Bremsstrahlung of 4FGL~J0330.7+5845e, fitted with an exponentially cut-off power-law (ECPL) proton/electron spectrum.}
    \label{brems+pp}
\end{figure}

Similarly, for protons we compared the same three spectral forms: PL, LP, and ECPL. Based on the BIC values shown in Tab.~\ref{diffrent model}, the ECPL spectrum was selected, described by Eq.~(\ref{ECPL}), where $E_0 = 1.7 \ \text{GeV}$, the proton spectral index is $\alpha = 1.8_{-0.5}^{+0.4}$, 
$E_{\text{cutoff}} = 17.1_{-7.3}^{+15.6} \ \text{GeV}$. The total proton energy above $3 \ \text{GeV}$ is $W_p({>}3\ \text{GeV}) = 2.1_{-0.2}^{+0.2} \times 10^{46} \ \text{erg}$, and above $1 \ \text{GeV}$ it is $W_p({>}1\ \text{GeV}) = 3.3_{-0.5}^{+0.6} \times 10^{46} \ \text{erg}$; the corresponding maximum log-likelihood value for this fit is $-0.86$. Likewise, the relatively large uncertainties in the spectral index and $E_{\rm cutoff}$ are likely due to the limited number of high-energy photons from the source, which weakly constrain the model. Nevertheless, the total proton energy $W_p$ remains sufficiently well-determined to perform the subsequent energy analysis. The fitted SED result for pion decay of 4FGL~J0330.7+5845e is shown in Fig.~\ref{brems+pp}.

\begin{table}
\centering
\caption{BIC values for different models.}
\label{diffrent model}
\begin{tabular}{lccc}
\hline
         & PL & LP & ECPL \\
\hline
Bremsstrahlung & 15.50 & 9.12 & 6.85 \\
PionDecay & 11.16 & 9.64 & 7.57 \\
\hline
\end{tabular}
\end{table}

\section{DISCUSSION}
\label{discussion}
We searched for potential $\gamma$-ray counterparts within a $0.5^\circ \times 0.5^\circ$ region centered on the ROI in several catalogs, including SNRcat~\citep{2012AdSpR..49.1313F}, Green's Galactic Supernova Remnant Catalog~\citep{2019JApA...40...36G}, the ATNF Pulsar Catalog~\citep{1998MNRAS.295..743L},  and the WISE AGN Catalog~\citep{2018ApJS..234...23A}. No plausible counterpart was identified, so based on positional argument, we consequently conclude that the source 4FGL~J0330.7+5845e is most likely associated with the SF region AFGL 490. In the following, we discuss the two particle acceleration mechanisms mentioned in Sect.~\ref{intro}.

\subsection{Stellar wind scenario}
We employ the following formula to estimate the stellar wind kinetic energy within the cluster core:
\begin{equation}
L_{w} = \frac{1}{2}\,\dot{M} v_{w}^{2}
= 1\times10^{35}\ \mathrm{erg\ s^{-1}}
\left( \frac{\dot{M}}{10^{-7}\ M_{\odot}\ \mathrm{yr^{-1}}} \right)
\left( \frac{v_{w}}{2000\ \mathrm{km\ s^{-1}}} \right)^{2}
\end{equation}
where \(\dot{M}\) is the mass-loss rate. The mass-loss rate is estimated using the relation \(\dot{M}_{\text{wind}} / \dot{M}_{\text{acc}} \sim 0.1\)~\citep{2000prpl.conf..759K}. 
(1) For the protostar AFGL 490, an accretion rate of 
\(\dot{M}_{\rm acc,protostar} \sim 10^{-5}\,M_\odot\,\text{yr}^{-1}\)~\citep{2021ApJ...912..108H} 
implies a wind mass-loss rate of 
\(\dot{M}_{\rm wind,protostar} \sim 10^{-6}\,M_\odot\,\text{yr}^{-1}\). 
As the protostar is expected to eventually evolve into a B-type main-sequence star, 
we adopt a typical B-type main-sequence stellar wind velocity 
of \(v_{w,\rm protostar} = 1000\,\text{km s}^{-1}\)~\citep{1990ApJ...361..607P}, 
which yields a kinetic luminosity of 
\(L_{w,\rm protostar} = 2.5 \times 10^{35}\,\text{erg s}^{-1}\). Assuming a protostellar age of \(\sim 10^5 \, \text{yr}\), the total integrated wind energy amounts to \(E_{\text{wind,protostar}} = 7.5 \times 10^{47} \, \text{erg}\).
(2) For the other YSOs in the core, we estimate a typical accretion rate of \(\dot{M}_{\text{acc,YSO}} \sim 10^{-8} \, M_\odot \, \text{yr}^{-1}\) per object~\citep{2006A&A...452..245N}, leading to a wind mass-loss rate of \(\dot{M}_{\text{wind,YSO}} \sim 10^{-9} \, M_\odot \, \text{yr}^{-1}\). We adopt a representative intrinsic wind velocity of \(v_{w,\text{YSO}} = 200 \, \text{km s}^{-1}\). This value is derived from the typical high-velocity component (HVC) of optical forbidden lines in T Tauri stars (\(\sim 100 \, \text{km s}^{-1}\); \citealt{1997A&AS..126..437H}), corrected by a factor of 2 to account for projection effects assuming random spatial orientations (i.e., a mean projection factor of \(\langle\cos i\rangle = 0.5\)). Consequently, the kinetic luminosity per YSO is \(L_{w,\text{YSO}} = 1 \times 10^{31} \, \text{erg s}^{-1}\). Multiplying by approximately 200 YSOs and an assumed age of \(10^6 \, \text{yr}\), the total wind energy from the low-mass population is \(E_{\text{wind,YSOs}} = 6 \times 10^{46} \, \text{erg}\).

Using the proton energy $W_p({>}3\ \text{GeV}) = 2.1 \times 10^{46}~\mathrm{erg}$ obtained from Sect.~\ref{hadronic}, we calculate the proton acceleration efficiency as $\eta_p = W_p({>}3\ \text{GeV}) / (E_{\text{wind,protostar}}+E_{\text{wind,YSOs}}) \approx 2.6\%$. However, an important point should be noted: if it is assumed that particles are accelerated by the protostar AFGL 490 and the other YSOs separately, the resulting acceleration efficiencies are approximately 2.8\% and 35\%, respectively. This clearly indicates that, under such a scenario, the stellar wind acceleration from the protostar AFGL 490 overwhelmingly dominates the process.

This derived efficiency appears problematic for several reasons. Firstly, as noted by \citet{2010MmSAI..81..181R}, the efficiency of particle acceleration by single stellar winds is expected to be very low because of various radiative and adiabatic losses. Furthermore, to date, no isolated massive star has been detected as a $\gamma$-ray source powered solely by its wind. Secondly, if we are to compare, the typical acceleration efficiency for supernova remnants (SNRs) is only around 10\%~\citep{1988ApJ...333L..65V}. Moreover, estimates for Young Massive Star Clusters (YMSCs) suggest that the efficiencies of particle acceleration by collective stellar winds are generally below 10\%, whilst in some cases it is even lower than 1\% \citep{2020SSRv..216...42B}. Therefore, it remains highly unlikely that the single protostellar wind serves as the primary mechanism for particle acceleration in this source.

\subsection{Jet scenario}
\cite{2021ApJ...912..108H} inferred from the Br-$\gamma$ emission line of the protostar AFGL 490 that its accretion luminosity is approximately $850\,L_\odot$. Then
by assuming a ratio of 0.075 between the jet kinetic luminosity and the accretion luminosity \citep{2007LNP...723...21C}, the kinetic luminosity of the protostar AFGL 490 jet is estimated to be $L_j = 2.44\times10^{35}\,\mathrm{erg\,s^{-1}}$.
Next,we estimate the time required to accelerate protons up to the necessary energy to produce the observed $\gamma$-ray emission by using the following equation:
\begin{equation}
t_{\mathrm{injection}} = \frac{W_p}{\eta L_{\mathrm{j}}} 
\end{equation}
where ${W_p}$ is the total proton energy(see Sect.~\ref{hadronic}), $\eta$ is the proton acceleration efficiency of the jet, which we vary between 2\% and 10\%~\citep{2021MNRAS.504.2405A}, and $L_{\mathrm{j}}$ is the jet kinetic luminosity. The resulting acceleration time ranges from $10^4$ to $10^5$ years. This range is consistent with the estimated age of the protostar AFGL 490 ($10^4$--$10^5$ years) and is also comparable to the estimated lifetime of $4 \times 10^4$ years for the HH 80-81 protostellar jet~\citep{2019ApJ...871..141Q}. Furthermore, this acceleration timescale falls within the same order of magnitude as the two cooling timescales ($t_{\mathrm{br}}$ and $t_{\mathrm{pp}}$) estimated in Sect.~\ref{origin}.

On the other hand, we also estimate the maximum charge energy achievable by the protostellar jet of AFGL 490 using the following relation \citep{2025ApJ...989L..25W}:

\begin{equation}
L_{\mathrm{j}} \geq 10^{38} \left( \frac{E_{\mathrm{max}}}{10\ \mathrm{PeV}} \right)^2 \widetilde{\omega} \beta^{-1} \sigma_{-1}^{-1} \ \mathrm{erg\ s^{-1}}
\end{equation}
where \( L_{\mathrm{j}} \) is the kinetic luminosity of the jet, \( E_{\mathrm{max}} \) is the maximum energy of charged particles, \( \widetilde{\omega} \) is a jet geometry factor assumed to be 1 for a collimated jet, and \( \beta \) is the maximum bulk velocity of the jet, for which we adopt a value of \( \beta \approx 0.0033 \) (corresponding to \( v = 1000\ \mathrm{km/s} \)). The parameter \( \sigma = B^2 / (2\pi \rho v^2) \) is the non-relativistic magnetization parameter, where \( \sigma_{-1} = \sigma / 0.1 \). For our calculation, we assume a mean magnetic field strength \( B = 0.1\ \mathrm{mG}\)~\citep{2019MNRAS.482.4687R,2025A&A...695A..11M} and a jet particle density \( n_{\mathrm{j}} = 10^{4}\ \mathrm{cm^{-3}} \)~\citep{2021MNRAS.504.2405A}, with \( v = \beta c \).

For a jet with a kinetic luminosity of \( L_j =2.44\times10^{35}\,\mathrm{erg\,s^{-1}} \), our calculation yields a maximum energy of charged particles of \( E_{\mathrm{max}} \approx 0.09 \mathrm{TeV} \). This result is consistent with the typical maximum proton energy of ~0.24 TeV predicted for acceleration in protostellar jets by \citet{2021MNRAS.504.2405A}. And for our ECPL proton/electron distribution model, we define the maximum proton/electron energy $E_{\mathrm{max}}$ as the energy containing 99.99\% of the total distribution. This yields $E_{p,\mathrm{max}} = 78.4\ \mathrm{GeV}$ and $E_{e,\mathrm{max}} = 23.6\ \mathrm{GeV}$. Both values lie comfortably within  the acceleration limit derived above.

\subsection{Diffusion of high-energy particles}
We use the following two expressions to calculate the diffusion radius of high-energy protons and electrons in the molecular cloud, respectively~\citep{2004vhec.book.....A}:
\begin{equation}
\label{pdiff}
R_{\mathrm{diff,p}}(E, t)
= 2 \sqrt{ D(E)\, t \,
\frac{ e^{t\delta/t_{\mathrm{pp}}} - 1 }{ t\delta / t_{\mathrm{pp}} } } ,
\end{equation}
and
\begin{equation}
\label{ediff}
R_{\mathrm{diff,e}}(E, t)
\simeq 2 \sqrt{ D(E)\, t \,
\frac{ 1 - \left( 1 - E/E_{\mathrm{cut}} \right)^{1-\delta} }
{ (1-\delta)\, E/E_{\mathrm{cut}} } } .
\end{equation}
Here the diffusion coefficient is defined as
$D(E)=D_{0}\left(E/10~\mathrm{GeV}\right)^{\delta}$, where $D_{0}$ typically lies in the range $10^{26}$--$10^{28}\ \mathrm{cm^{2}\,s^{-1}}$.
In this work, we adopt $D_{0}=10^{26}\ \mathrm{cm^{2}\,s^{-1}}$, which represents a denser environment, to obtain a lower limit on the diffusion radius, and set $\delta=0.5$.
The diffusion time is taken to be $t=10^{4}$--$10^{5}\ \mathrm{yr}$.

For Eq.~(\ref{pdiff}), we adopt the proton--proton interaction timescale
$t_{\mathrm{pp}} \simeq 1.41\times10^{5}\ \mathrm{yr}$,
as derived in Sect.~\ref{hadronic} and in Eq.~(\ref{ediff}), the time-dependent cutoff energy is given by
$E_{\mathrm{cut}}(t) \simeq m_{\mathrm{e}}c^{2}/\left[(7\times10^{-20}\ \mathrm{eV\,cm^{-3}\,s^{-1}})\,t\right]$.
The resulting diffusion radius for different particle energies are listed in
Tab.~\ref{diff table}. Considering that the characteristic size of the molecular cloud derived in
Sect.~\ref{density} is $R \sim 6.5\ \mathrm{pc}$,
and that the physical scale of the 4FGL~J0330.7+5845e is $\sim 8.2\ \mathrm{pc}$,
as derived from the $ext = 0.47^\circ$, it is therefore plausible that these high-energy particles can diffuse within
the cloud and produce the observed extended $\gamma$-ray source, 4FGL~J0330.7+5845e.

\begin{table}
\centering
\caption{The ranges of diffusion radius (in units of pc) for particles with different energies, for diffusion times ranging from $10^{4}$ to $10^{5}\ \mathrm{yr}$.}
\label{diff table}
\begin{tabular}{lccc}
\hline
Energy & 3GeV & 10GeV & 78.7GeV \\
$R_{\mathrm{diff,p}}$(pc) & 2.7 - 9.3 & 3.7 - 12.6 &  6.2 - 21.1\\
\hline
Energy & 3GeV & 10GeV & 21.2GeV \\
$R_{\mathrm{diff,e}}$(pc) & 2.7 - 8.5 & 3.6 - 11.5 & 4.4 - 13.9 \\
\hline
\end{tabular}
\end{table}

\section{CONCLUSION}
\label{conclusion}
In this paper, we report the detection of GeV $\gamma$-ray emission toward the star-forming region AFGL 490. Through a relocalization of the source 4FGL J0330.7+5845, we have identified a significantly extended source, which is designated 4FGL~J0330.7+5845e. The emission is well-modeled by a Gaussian spatial template with an extension radius of $0.47^\circ$ and a high test statistic (TS $= 172$). The centroid of this extended emission is located $0.144^\circ$ from the protostar AFGL 490, a separation that lies well within the $95\%$ containment radius ($r_{95} = 0.147^\circ$). The source exhibits significant spatial extension across multiple energy ranges, including 0.3–300 GeV, 0.3–1 GeV, and 1–300 GeV, with TS$_{\rm ext} > 16$ in all cases, and its spectral energy distribution is fit by a log-parabolic function with notable curvature, indicative of a high-energy cutoff ($\alpha = 2.312\pm 0.112$, $\beta = 0.234\pm 0.089$, $E_b = 1038$ MeV).

Our investigation, which combines an analysis of the ambient gas distribution with theoretical modeling, does not allow us to unambiguously determine whether the leptonic or hadronic channel dominates the observed emission. In the absence of compelling alternative candidates, however, the protostellar jet of AFGL 490 remains a plausible acceleration scenario capable of accounting for the observed gamma-ray emission.

Future observations with next-generation telescopes, offering superior angular resolution, will be pivotal in revealing any potential fine-scale structure or additional point sources within this complex region. Furthermore, multi-wavelength campaigns, combining data across different bands, will be essential to place more stringent constraints on the physical parameters of the protostellar jet and to further elucidate the particle acceleration processes at work.

\section*{Acknowledgements}
This work is supported by the National Natural Science Foundation of China (NSFC) grant 12273122, National Astronomical Data Center, the Greater Bay Area, under grant No. 2024B1212080003, and science research grant from the China Manned Space Project under CMS-CSST-2025-A13. This research made use of the data from the Milky Way Imaging Scroll Painting (MWISP) project, which is a multi-line survey in $^{12}$CO/$^{13}$CO/C$^{18}$O along the northern galactic plane with PMO-13.7m telescope. We are grateful to all the members of the MWISP working group, particularly the staff members at PMO-13.7m telescope, for their long-term support. MWISP was sponsored by National Key R\&D Program of China with grants 2023YFA1608000 \& 2017YFA0402701 and by CAS Key Research Program of Frontier Sciences with grant QYZDJ-SSW-SLH047.

\section*{Data Availability}
The Fermi-LAT data used in this work are publicly available, and are provided online at the NASA-GSFC Fermi Science Support Center\footnote{\url{https://fermi.gsfc.nasa.gov/ssc/data/access/lat/}}. We made use of the CO data from the MWISP\footnote{\url{http://www.dlh.pmo.cas.cn/ENGLISH/MWISP/}} to study the distribution of \ce{H2}, 
and the \ce{HII} and \ce{HI} data are taken from Planck legacy archive\footnote{\url{https://pla.esac.esa.int/pla/\#home}} and HI4PI\footnote{\url{https://cdsarc.u-strasbg.fr/viz-bin/qcat?J/A+A/594/A116}}, respectively.



\bibliographystyle{mnras}
\bibliography{example} 





\bsp	
\label{lastpage}
\end{document}